\documentclass[conference]{IEEEtran}
\IEEEoverridecommandlockouts
\usepackage{cite}
\usepackage{amsmath,amssymb,amsfonts}
\usepackage{algorithmic}
\usepackage{graphicx}
\usepackage{textcomp}
\DeclareMathOperator*{\argmin}{arg\,min}
\usepackage{xcolor}
\usepackage{subcaption}
\usepackage{tabularx}
\usepackage{multirow}
\usepackage{booktabs}
\usepackage{hyperref}
\usepackage[capitalise]{cleveref}
\def\BibTeX{{\rm B\kern-.05em{\sc i\kern-.025em b}\kern-.08em
    T\kern-.1667em\lower.7ex\hbox{E}\kern-.125emX}}

\renewcommand{\baselinestretch}{0.948}
    
\begin{document}

% %\title{Bayesian Physics-informed Neural Networks for System Identification of Power System Dynamics with Uncertainty\\
% \title{System identification of Power System Dynamics under Noise using Bayesian Physics-informed Neural Networks and Variational Inference 
% %\thanks{Identify applicable funding agency here. If none, delete this.}
% }

% \title{Bayesian Physics-Informed Neural Networks for System Identification of Noisy Power System Dynamics 
% %\thanks{Identify applicable funding agency here. If none, delete this.}
% }

\title{Bayesian Physics-Informed Neural Networks for Robust System Identification of Power Systems 
}
\author{\IEEEauthorblockN{Simon Stock$^1$, Jochen Stiasny$^2$, Davood Babazadeh$^1$, Christian Becker$^1$, Spyros Chatzivasileiadis$^2$}
\IEEEauthorblockA{\textit{$^1$Institute of Power and Energy Technology, Hamburg University of Technology,} Hamburg, Germany \\
$^2$\textit{Department of Wind and Energy Systems, Technical University of Denmark,} Kgs. Lyngby, Denmark \\
simon.stock@tuhh.de, jbest@dtu.dk, davood.babazadeh@tuhh.de, c.becker@tuhh.de, spchatz@dtu.dk}
}

\maketitle

\begin{abstract}
This paper introduces for the first time, to the best of our knowledge, the Bayesian Physics-Informed Neural Networks for applications in power systems. Bayesian Physics-Informed Neural Networks (BPINNs) combine the advantages of Physics-Informed Neural Networks (PINNs), being robust to noise and missing data, with Bayesian modeling, delivering a confidence measure for their output. Such a confidence measure can be very
valuable for the operation of safety critical systems, such as
power systems, as it offers a degree of `trustworthiness' for the neural network output. This paper applies the BPINNs for robust identification of the system inertia and damping, using a single machine infinite bus system as the guiding example. The goal of this paper is to introduce the concept and explore the strengths and weaknesses of BPINNs compared to existing methods. We compare BPINNs with the PINNs and the recently popular method for system identification, SINDy. We find that BPINNs and PINNs are robust against all noise levels, delivering estimates of the system inertia and damping with
significantly lower error compared to SINDy, especially as the
noise levels increases.
%
% In recent years the dynamics of power systems have changed due to the decoupling of conventional power plants and the introduction of inverter-based generation and loads. This leads to the necessity of system identification possibilities to enable system awareness and therewith keep the grid stable. Phasor measurement units provide the infrastructure to enable fast system state assessment.   
% %However, two problems can be found in the present system identification algorithms that will be addressed in this paper. First, many approaches are prone to noisy data, which appears to be present in standard measurement systems. Second, it is not possible to see the quality of the algorithms estimate without knowing the target value. 
% Based on these measurements, a Bayesian physics-informed neural network is proposed in this paper that combines the advantages of Bayesian neural networks, namely being robust against noisy data and quantifying the uncertainty about their outputs through the standard deviation and the ability of physics-informed neural networks to estimate the parameters of an underlying physical system without requiring access to the target values in the training phase. The proposed approach is evaluated using a single-machine-infinite-bus system investigating noise sensitivity and is compared to a state of the art algorithm for sparse identification of nonlinear dynamics (SINDy) and the physics-informed neural network (PINN). 
\end{abstract}

\begin{IEEEkeywords}
Bayesian Physics-Informed Neural Networks, System Identification, Swing Equation, Power System Dynamics
\end{IEEEkeywords}

\section{Introduction}
Power systems dynamic behavior is drastically changing due to the rapid integration of converter-connected energy sources, and the gradual decommissioning of conventional power plants. This leads to overall lower levels of inertia and a less damped system. 
%which inherently provided stabilizing energy through their inertia and damping. 
% This does not hold for the inverter-based resources that has been introduced in the last years, e.g. wind power plants, PV. 
Additionally, the uncertain and fluctuating output of renewable energy sources, such as wind farms and solar PVs, results in frequent changes of the power system dynamic characteristics. This requires advanced approaches to assess the power system state rapidly, continuously, and with confidence, in order to determine the current level of inertia and damping and commit accordingly the necessary frequency reserves to avoid instability and blackouts.
% makes assessing the power system dynamics mandatory to keep track of the system stability in situations of high volatility. 
The proliferation of different types of sensors, such as phasor-measurement units (PMUs), smart meters, and others, has enabled the high-frequency monitoring of power system behavior through the collection of significant amounts of data. Such data can help towards the so-called online system identification, for which multiple approaches exist in the literature as we attempt to summarize below. 
%Usually, the system to be determined is formulated as a state-space representation, so the derivative of the measured states has to be found. 
To determine the system dynamics, the temporal derivative of the data is usually calculated. This helps estimate the parameters of a set of differential equations which describe the system behavior.
% increasing number of measurements, the high-frequency tracking of power system behavior has been enabled. So, multiple approaches that utilize these data for system identification of dynamics can be found in the literature.

System identification has so far used classical linear approaches such as Dynamic Mode Decomposition (DMD) \cite{Yang2021_dmd} or Koopman theory \cite{Susuki2018}, and approaches that rely on filtering methods such as Kalman filtering \cite{zhao2019}; recently, there has been increased focus on parsimonious approaches, such as SINDy \cite{Brunton2016} to cope with nonlinearities. We briefly detail SINDy in \cref{sec:SINDy} of this paper. Nevertheless, SINDy is prone to noise in the data due to its point-wise differentiation that could lead to large errors. In that respect, Physics-Informed Neural Networks have appeared as a very promising alternative \cite{Stiasny2020}, as they fit the whole trajectory before differentiation; this helps with smoothing the values before getting differentiated, which makes them robust against noise. For example, \cite{Stiasny2020} shows that under strong nonlinearities and noise, Physics-Informed Neural Networks (PINN) can estimate the inertia and damping of the system with small error.
%On the other hand, Machine Learning has appeared as a very promising alternative to cope with unobsverability \cite{Mestav2019} and nonlinearities \cite{Stiasny2020} appearing in power systems. For example, \cite{Stiasny2020} shows that with increasing nonlinearities and shorter time scales appearing from the integration of converter-connected resources, Physics-Informed Neural Networks (PINN) can predict the inertia and damping of the system with lower error compared to an Unscented Kalman filter. 

However, SINDy and PINNs are by design not able to give a statement about the confidence level of their estimates. In contrast, Bayesian approaches have the capability to quantify uncertainty as they try to fit a whole family of functions instead of a single function. By inferring a posterior distribution of the parameters from the measurement data, the mean value can be seen as the best guess of the estimate and the standard deviation as a measure of its uncertainty. Following this path, \cite{Petra2017} has presented an approach for system identification of power system dynamics utilizing Bayesian inference. However, this approach requires prior knowledge about the region of the target parameter.

To the best of our knowledge, this paper introduces for the first time the Bayesian Physics-Informed Neural Networks for use in power system applications. Bayesian Physics-Informed Neural Networks (BPINNs)
% So, to solve this issue we apply the Bayesian Physics-informed Neural Networks (BPINN) to system identification of power system dynamics and compare it to the PINN and the SINDy algorithm. This approach
combine the advantages of the PINNs, being robust to noise and missing data, with Bayesian modelling that is able to deliver a confidence measure of the estimate, through the standard deviation of the posterior. In contrast with existing approaches, BPINN does not require prior knowledge about either the target values (parameters) and their potential range or the noise in the measurement data, i.e. fidelity of the sensors \cite{YANG2021}. 

BPINNs can have a wide range of applications in power systems, going well beyond robust system identification that we explore in this paper. The goal of this paper is to introduce the concept and explore the strengths and weaknesses of BPINNs compared to standard PINNs and the powerful conventional approaches. Enriching, however, Physics-Informed Neural Networks, with a confidence measure about their estimate -- as Bayesian PINNs do -- is expected to open a wide range of opportunities for application in power systems. 

This paper is structured as follows. \Cref{chapter:methodology} formulates the problem, provides the related equations, and presents the proposed approach in detail. \Cref{chapter:results} investigates the BPINN and compares its performance to SINDy and the PINN. \Cref{chapter: conclusion} concludes.
%It can also be applied in a distributed manner to determine the system dynamic parameters at different points in the grid, so we first evaluate the performance utilizing the single-machine-infinite-bus system which can be seen as an aggregated model of a certain part of the power system. Furthermore, it also gives a good understanding of the sensitivities of the proposed approach.
%\section{Problem formulation}
%\label{chapter: problem_formulation}
\section{Methodology}
\label{chapter:methodology}
In this section, we introduce the Bayesian Physics-Informed Neural Network (BPINN) formulation used throughout this paper. To do that, first, we briefly revisit the neural network and PINN formulations, and then we present the Bayesian PINN and highlight similarities between the deterministic and Bayesian formulations.

Assume a dynamic model described by the following differential equations (in Section~\ref{sec:Method_Simulation}, we detail the formulation for the specific power system dynamics problem): 
\begin{align}
    \Dot{\boldsymbol{x}}&=f(\boldsymbol{x}, \boldsymbol{u};\boldsymbol{\lambda})
\end{align}
with solution $\boldsymbol{x}(t,\boldsymbol{u})$, $\boldsymbol{x}$ representing the states of the system and $\boldsymbol{u}$ the inputs of the system. $\boldsymbol{\lambda}$ describes the system parameters that are to be determined and $f$ is the operator mapping the system parameters to the states.
%The formulated problem is based on a dynamic model that can be described by the following differential algebraic equations 
%\begin{align}
%    \Dot{\boldsymbol{x}}&=\boldsymbol{f(x,y,u;\lambda)}\\
%    \boldsymbol{0}&=\boldsymbol{g(x,y,u;\lambda)}
%\end{align}
%with $\boldsymbol{x}$ representing the states of the system, $\boldsymbol{y}$ the algebraic variables, $\boldsymbol{u}$ the inputs of the system and $\boldsymbol{\lambda}$ the system parameters that are to be determined.\\
%For power system frequency dynamics this formulation is based on the swing equation and can be written as follows for a single generator of second order with no model losses.
%\begin{align}
%    \label{eq:swing_equation}
%    m_k\ddot{\delta}+d_k\Dot{\delta}+\sum_j B_{kj} V_k V_j sin(\delta_k-\delta_j)-P_k=0
%\end{align}
%Therein, $m_k$ denotes the inertia constant of the generator, $d_k$ the damping coefficient, $B_{kj}$ is the entry of the susceptance matrix, $V_k$ and $V_j$ the voltages of the $j$th and $k$th bus and $P_k$ the mechanical power. $\delta_k$ and $\delta_j$ represent the voltage angles at the corresponding buses. \\
%The aim of this paper is to determine the parameters $\boldsymbol{\lambda}$, namely $m_k$ and $d_k$ of the system based on measurement data while providing confidence about the estimate. Therefore, a measurement trajectory of $\delta_k$ and $\dot{\delta_k}$ as well as the power $P_k$ is utilized.
Neural networks are generally used as function approximators for various problems. In the case of 
%non-linear
dynamic systems, their goal is to determine a surrogate model ${g}(t)$ that maps a time-dependent input vector to the target trajectory of the states $\boldsymbol{x}$ as follows:
\begin{align}
    g(t; \boldsymbol{\Theta}) =\hat{\boldsymbol{x}}(t; \boldsymbol{\Theta})\approx \boldsymbol{x}(t, \boldsymbol{u}; \boldsymbol{x_0}, \boldsymbol{\lambda})
\end{align}
Here, $\boldsymbol{x_0}$ describes the initial state of the system.
%with parameters $\boldsymbol{\lambda}$. 
The surrogate model should be determined using a set of measurement data $\mathcal{D}$ with $\mathcal{D}=\{ (\boldsymbol{x}^{(i)}, \boldsymbol{u}^{(i)}) \}_{i=1}^N$. 
%To illustrate the idea of PINNs, their loss formulations will be revisited in the following.
During training, the distance between $\hat{\boldsymbol{x}}(t)$ and the target trajectory is minimized by varying the neural network parameters $\boldsymbol{\Theta}$, i.e. the NN weights and biases. Equation \eqref{eq:forward_optimization_NN} presents the formulation of such an optimization problem, using the root-mean-squared error as the distance measure. The training data $\boldsymbol{x} \in \mathcal{D}$ follow an additive Gaussian noise model $\boldsymbol{\eta}$ with unknown standard deviation. Such a Neural Network training procedure is a supervised learning problem.
\begin{align}
    \min_{\Theta}  \frac{1}{N} \sum_{i=1}^N\sqrt{(\hat{\boldsymbol{x}}^{(i)}(\Theta)-(\boldsymbol{x}_{true}^{(i)}+\boldsymbol{\eta}^{(i)}))^2}
    \label{eq:forward_optimization_NN}
\end{align}
 
%This modeling approach is generally physics-agnostic and might require large amounts of data to achieve a good fit. Therefore, 
Physics-Informed Neural Networks (PINNs) have been proposed to determine the parameters $\boldsymbol{\lambda}$ \cite{RAISSI2019} utilizing automatic differentiation to find $\frac{d}{dt}\hat{\boldsymbol{x}}$ and feed this into a physics regularization term: 
%considering the previously introduced swing equation.
\begin{align}
    %h(t,\boldsymbol{u})&=\dot{\boldsymbol{x}}-f(\boldsymbol{x};\boldsymbol{\lambda}) &\approx 0\\
    h(t,\boldsymbol{u}; \boldsymbol{\Theta}, \boldsymbol{\lambda})&=\frac{d}{dt}\hat{\boldsymbol{x}}-f(\hat{\boldsymbol{x}}, \boldsymbol{u};\boldsymbol{\lambda}) \stackrel{!}{=} 0
    %h(t,\boldsymbol{u}, \Theta, \lambda)&=\frac{d}{dt}\hat{\boldsymbol{x}}(t,\Theta)-f(\hat{\boldsymbol{x}}(t,\Theta), \boldsymbol{u};\boldsymbol{\lambda}) &\approx 0
    %\frac{d}{dt}g(t, \Theta)-f(g(t, \Theta);\boldsymbol{\lambda}) \approx 0
\label{eq:residual_loss}
\end{align} 
%\begin{align}
%    \hat{r_k}(t)=\hat{m_k}\ddot{\delta}+\hat{d_k}\Dot{\delta}+\sum_j a_{kj} V_k V_jsin(\delta_k-\delta_j)-P_k\approx 0
%\end{align}
The goal is to determine the parameters $\boldsymbol{\lambda}$ of a known function $f$.
%The goal is to determine a surrogate function $g$ that fits a more complex known function $f$. 
This, in theory, does not require any external data, and, therefore makes the problem an unsupervised learning problem, as shown in \cite{Stiasny2020}. 
%However, considering that the known function $f$ is often inaccurate or contains unknown parameters that we need to determine, as in our case, 
The Physics-Informed Neural Networks usually combine \eqref{eq:forward_optimization_NN} and \eqref{eq:residual_loss} in the loss function of the neural network training, 
to determine the system states $\boldsymbol{x}$ and parameters $\boldsymbol{\lambda}$. 
%When it comes to System Identification, as is the focus of this paper, PINNs use the physics regularization term as a guide to help them accurately determine the unknown parameters, based on the available training data they receive.

\subsection{Bayesian PINN}
\begin{figure}[h!]
    \centering
    \includegraphics[clip,width=3.5in]{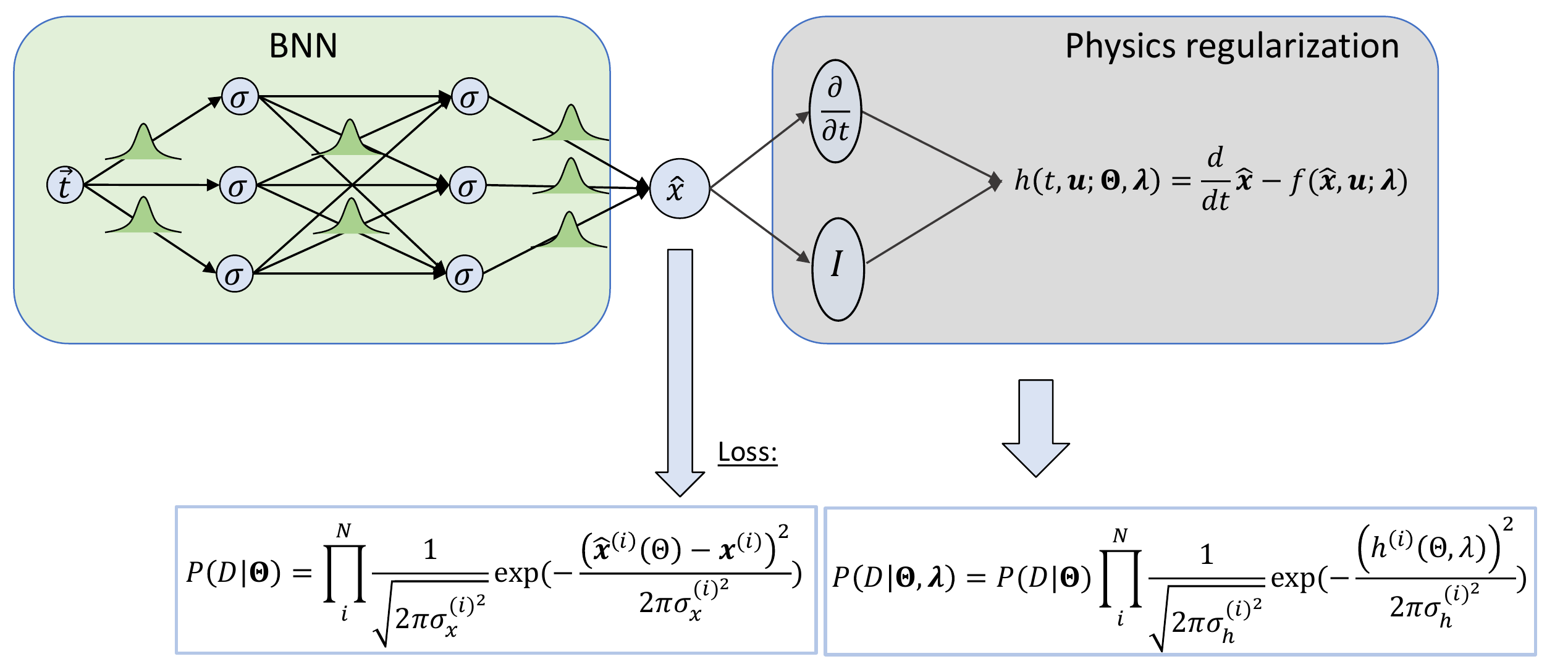}
    \caption{Bayesian Physics-Informed Neural Network schematic}
    \label{fig:BPINN_structure}
\end{figure}

Bayesian Neural Networks (BNNs) were introduced about three decades ago to address the problem of uncertainty in data, and were shown to deliver good estimates in the presence of noise \cite{Kononenko1989}. In BNNs, the weights, biases and outputs of the neural network are formulated as probability distributions and during the training phase, Bayes' theorem is used to infer their posterior distributions $P(\boldsymbol{\Theta}|\mathcal{D})$ based on the data $\mathcal{D}$ and the specified prior $P(\boldsymbol{\Theta})$ of the parameters. As a result, the Bayesian neural network can be seen not as a single function, but as a family of functions that also deliver a confidence measure about their function approximation
%, i.e. a mean of the posterior, 
based on the variance of the posterior.
This formulation also leads to a different loss formulation compared to \eqref{eq:forward_optimization_NN} which is now based on the following equation. 
\begin{align}
    P(\mathcal{D}|\boldsymbol{\Theta})=\prod_i^N\frac{1}{\sqrt{2 \pi \sigma_x^{(i)^2}}}\text{exp}(-\frac{(\hat{\boldsymbol{x}}^{(i)}(\boldsymbol{\Theta})-\boldsymbol{x}^{(i)})^2}{2 \sigma_x^{(i)^2}})
     \label{eq:BNN_loss}
\end{align}
To keep the approach widely applicable, here we do not assume the measurement error to be known a priori; this means that $\sigma_x$ is also a parameter that has to be determined during training. 
%of a temporal sequence with measurement noise normally distributed around the hidden real value like $\boldsymbol{h}(t)=\boldsymbol{h}_{true}(t)+\boldsymbol{\epsilon}$. 
The posterior distribution of parameters $\boldsymbol{\Theta}$ can then be calculated based on 
\begin{align}
    P(\boldsymbol{\Theta|}\mathcal{D})=\frac{P(\mathcal{D}\boldsymbol{|\Theta}) P(\boldsymbol{\Theta})}{P(\mathcal{D})}
\end{align}
using a Bayesian inference algorithm.
Still, this formulation is only applicable to forward problems where the goal is to approximate the temporal sequence of states $\boldsymbol{x}(t,\boldsymbol{u})$. In a power systems context, this allows us to estimate e.g. the evolution of frequency or rotor angle, but it does not allow us to perform a system identification.
%, e.g. estimate the inertia of the system (which is here included in the set $\boldsymbol{\lambda}$). 
To determine the parameters $\boldsymbol{\lambda}$, we need to introduce a physical regularization similar to \eqref{eq:residual_loss}, as proposed in \cite{YANG2021} for Bayesian approaches, introducing the so-called Bayesian Physics-Informed Neural Network (BPINN). By doing that, not only has the trajectory estimate from the BNN to fit the true trajectory, but also has to validate the physical equation. As a result, the loss function for the BPINN training becomes [notice \eqref{eq:BPINN_loss} includes \eqref{eq:BNN_loss}]: 
\begin{align}
    P(\mathcal{D}| \boldsymbol{\Theta}, \boldsymbol{\lambda})= P(\mathcal{D}|\boldsymbol{\Theta}) \prod_i^N\frac{1}{\sqrt{2 \pi \sigma_h^{(i)^2}}}\text{exp}(-\frac{(h^{(i)}(\boldsymbol{\Theta}, \boldsymbol{\lambda}))^2 }{2 \sigma_h^{(i)^2}})
    \label{eq:BPINN_loss}
\end{align}
The joint posterior of $\boldsymbol{\Theta}$ and $\boldsymbol{\lambda}$ can be determined following Bayes' theorem based on the prior distributions of $P(\boldsymbol{\Theta)}$ and $P(\boldsymbol{\lambda)}$. 
\begin{align}
    P(\boldsymbol{\Theta, \lambda|}\mathcal{D})=\frac{P(\mathcal{D}\boldsymbol{|\Theta, \lambda}) P(\boldsymbol{\Theta, \lambda})}{P(\mathcal{D})}
\label{eq:posterior_inverse}
\end{align}
This process of a Bayesian Physics-Informed Neural Network is illustrated in \cref{fig:BPINN_structure}.

%\begin{figure*}[h!]
%    \centering
%    \includegraphics%[clip,width=3.5in]
%    [height=6cm]{BPINN_structure.pdf}
%    \caption{BPINN structure}
%    \label{fig:BPINN_structure}
%\end{figure*}
%\begin{figure}[h!]
%    \centering
%    \includegraphics[clip,width=3.5in]{BPINN_structure2.pdf}
%    \caption{BPINN structure}
%    \label{fig:BPINN_structure}
%\end{figure}
However, the posterior distribution of unknown parameters $P(\boldsymbol{\Theta,\lambda|}\mathcal{D})$ is intractable. To find it, we use variational inference, which formulates this problem as an optimization, so computationally efficient optimization libraries can be used. 
%We approximate $P(\boldsymbol{\Theta,\lambda|}\mathcal{D})$ by a family of density functions $Q(\boldsymbol{\Theta, \lambda;} \,\xi)$. $\xi$ is now tuned to minimize the Kullback-Leibler divergence \cite{Joyce2011}, which gives us the member of $Q$ which is closest to $P$:
%\begin{align}
%    \begin{split}
%        \min_{\xi} D_{KL}(Q(\boldsymbol{\Theta, \lambda;} \,\xi)&||P(\boldsymbol{\Theta,\lambda|}\mathcal{D})) \simeq \mathbb{E}_{\boldsymbol{\Theta},\boldsymbol{\lambda} \sim Q}[ln\, Q(\boldsymbol{\Theta, \lambda;} \,\xi)\\ 
%        &- ln \,P(\boldsymbol{\Theta}, \boldsymbol{\lambda}) - ln \,P(\mathcal{D}|\boldsymbol{\Theta}, \boldsymbol{\lambda})]
%    \end{split}
%\end{align}
%So, $Q$ can be used as a proxy for conditional density $P$.
To solve this optimization problem, the Stein Variational Gradient Descend (SVGD) algorithm is used in this paper as proposed in \cite{LIU2016}.

\subsection{Simulation}
\label{sec:Method_Simulation}
\begin{figure}[h]
    \centering
    \includegraphics[clip, width=3.5in]{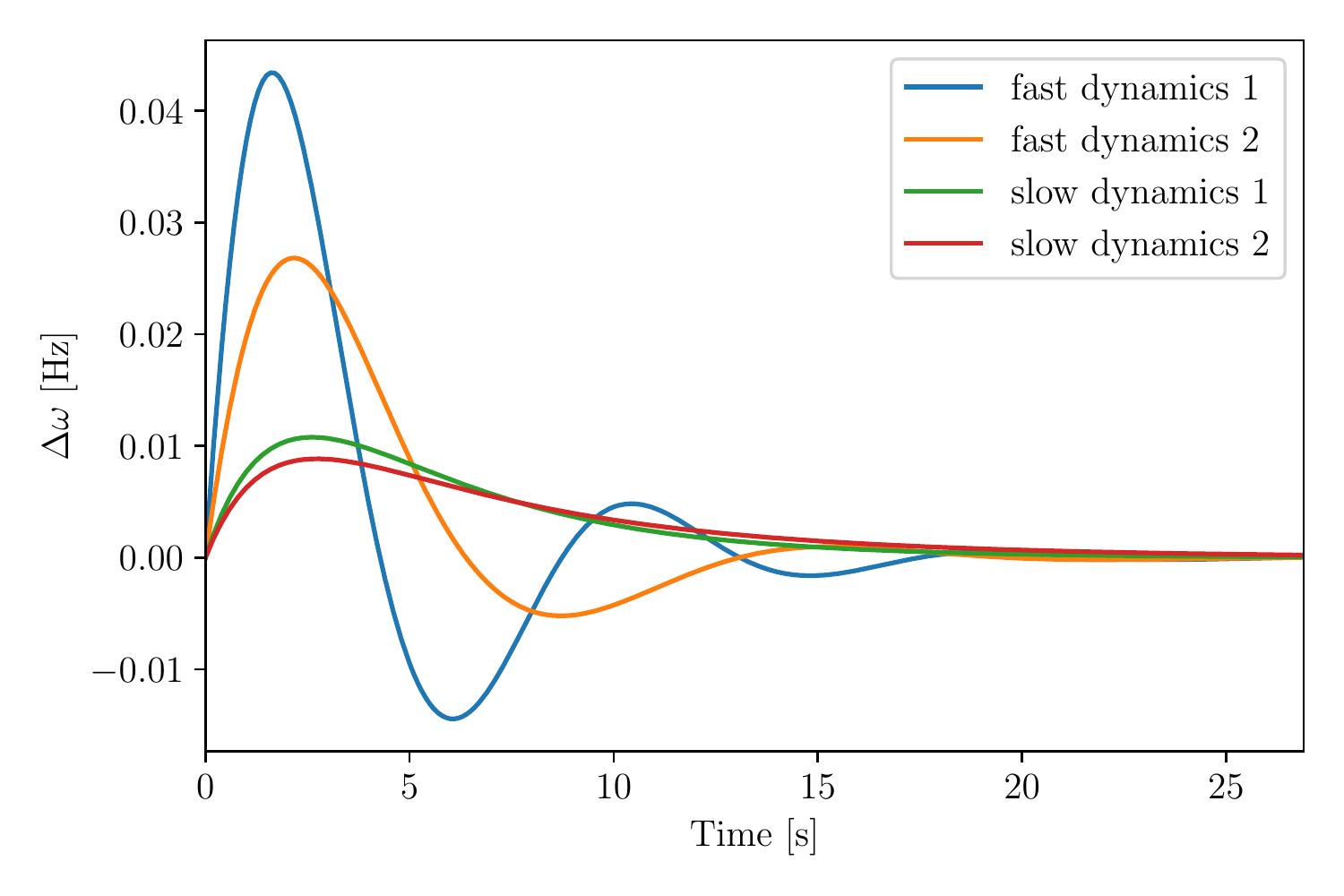}
    \caption{$\omega$ and $\delta$ trajectories of all scenarios ($0\,\%$ noise)}
    \label{fig:trajectories_of_scenarios}
\end{figure}

To help our understanding about the BPINN performance, we introduce the Bayesian Physics-Informed Neural Networks for power systems using a single-machine infinite-bus system, as follows: 
\begin{align}
\dot{\delta} &= \Delta \omega \label{eq:swing_equation2}\\
\dot{\Delta\omega} &= \frac{1}{m}(P - d\, \Delta\omega - B \, \text{sin}\delta)
\label{eq:swing_equation}
\end{align}
with $m$ being the inertia, $d$ the damping, $B$ the entry of the susceptance matrix and $P$ the active power. The states of the system are the angle and the frequency deviation: $\boldsymbol{x}=\{\delta, \Delta\omega\}$. A single-machine infinite bus system can also often represent an aggregate of a larger power system; estimating the parameters $\boldsymbol{\lambda}=\{ m,d \}$ through the BPINN delivers an estimate for the system inertia and damping, including a confidence measure about the estimate.
%internal.
% With this simple system, we are able to investigate the behavior of the BPINN in detail. 
% Furthermore, when estimating the parameters $\boldsymbol{\lambda}=\{ m,d \}$ at one point in the power system, it can be assumed that the system of interest is aggregated as a single generator. The power $P$, the angle $\delta$ and the frequency deviation $\Delta \omega$ are measured to be used in the estimation of parameters $m$ and $d$.
%The previously described temporal sequence to be estimated by the BPINN maps to $ \boldsymbol{x(t,u)}= \{\delta(t,u), \Delta \omega(t,u)\}$ in power system dynamics. 
We studied four different scenarios ranging from fast system dynamics to slow system dynamics as described in \cref{tab:scenarios}. The corresponding trajectories, assuming no measurement noise, are shown in \cref{fig:trajectories_of_scenarios}. For all scenarios, $B=0.2$ and $P=0.1$, while the initial states are $\boldsymbol{x}_0=\{0,0\}$.  
\begin{table}[!th]
\renewcommand{\arraystretch}{1.2}
\caption{Evaluation scenarios}
\label{tab:scenarios}
\centering
\begin{tabular}{lcc}
\toprule
\textbf{Scenario} & $m$ in p.u. & $d$ in p.u.\\
\midrule
fast dynamics 1 (fd1) & $0.3$ & $0.15$ \\
fast dynamics 2 (fd2) & $0.6$ & $0.3$ \\
slow dynamics 1 (sd1) & $1.4$ & $1.1$  \\
slow dynamics 2 (sd2) & $1.7$ & $1.4$ \\
\bottomrule
\end{tabular}
\end{table}

For all tests, a BPINN with 10 neurons and 1 hidden layer, and a standard trajectory length of $T= 27\, s$ was used at a sampling frequency of $10\,\text{Hz}$. The inputs of the BPINN are $t$ and $P$. 
%In \cref{fig:BPINN_distribution} the results presented in the following section are visualized for 10 runs of the same trajectory. The bar in the middle described the mean estimate, best guess, while the box lower bound of the box is the $25\,\%$ percentile and the upper the $75\,\%$ percentile, these bounds are also presented in the table \cref{tbl:Noise_results} and \cref{tbl:trajectory_noise_resuts} together with the mean estimate. It can be seen from \cref{fig:BPINN_distribution} that the estimates do not vary significantly over 10 runs, so only one run is performed for each of the following investigation scenarios. 
To avoid an initialization bias, 10 runs are performed for all tests and the average mean and average standard deviation are used for the BPINN results presented in \cref{tbl:Noise_results} and \cref{tbl:trajectory_noise_resuts}. 
For SINDy and the PINN 10 runs are performed per scenario and the average of the estimates is presented in the tables \cref{tbl:Noise_results}, \cref{tbl:trajectory_noise_resuts}. In contrast to BPINNs, which also output the standard deviation of their estimate in each run, notice that SINDy and PINN do not offer any confidence measure, they only offer a point estimate.
The BPINN was implemented in Python using packages $pytorch$ and $numpyro$; the code can be found in GitHub \cite{Git}.

%\begin{figure}
%    \renewcommand{\arraystretch}{0.3}
%    \centering
%    \begin{subfigure}{0.48\linewidth}
%        \includegraphics[width=\linewidth]{d_boxplot_multiple_runs.pdf}
%        \caption{BPINN estimate $d$ over multiple runs}
%        \label{fig:BPINN_dist_d}
%    \end{subfigure}
%    \hfill
%    \begin{subfigure}{0.48\linewidth}
%        \includegraphics[width=\linewidth]{m_boxplot_multiple_runs.pdf}
%        \caption{BPINN estimate $m$ over multiple runs}
%        \label{fig:BPINN_dist_m}
%    \end{subfigure}
%    \caption{BPINN error distributions}
%    \label{fig:BPINN_distribution}
%\end{figure}

\subsection{SINDy algorithm}
\label{sec:SINDy}
We compare our approach with one of the recently popular approaches for system identification: the \textbf{S}parse \textbf{I}dentification of \textbf{N}onlinear \textbf{Dy}namics (SINDy) algorithm, proposed in \cite{Brunton2016}. This approach is part of the parsimonious approaches which focus on the active terms of the dynamic systems, to reduce the computational effort and produce explainable results. To determine a generalized model with as few active terms as possible, sparse regression is utilized, penalizing the number of active terms in the dynamic model following the equation 
\begin{align}
    \argmin_{\Xi}||\dot{\boldsymbol{x}}-\boldsymbol{\zeta}(\boldsymbol{x})\boldsymbol{\Xi}||_2+\boldsymbol{\nu} \boldsymbol{||\Xi||_1} 
\end{align}
with $\dot{\boldsymbol{x}}$ being the derivatives of the states and $\boldsymbol{\zeta}(\boldsymbol{x})\boldsymbol{\Xi}$ a representation of the dynamic system with candidate nonlinear functions $\boldsymbol{\zeta}(\boldsymbol{x})$ and $\boldsymbol{\Xi}$ a coefficient vector of the active terms. The use of $\boldsymbol{\Xi}$ in the regularizing term $\boldsymbol{\nu}\boldsymbol{||\Xi||_1}$ accounts for a sparse model. However, when trying to determine the unknown parameters $\boldsymbol{\lambda}$ of a known set of equations, i.e. known set of $\boldsymbol{\zeta}(\boldsymbol{x})$, the regularizing $\boldsymbol{\nu}\boldsymbol{||\Xi||_1}$ can be neglected. For more information, the interested reader can refer to \cite{Brunton2016}.
In this paper, the PySINDy library was used \cite{Kaptanoglu2022}. It should be mentioned here that SINDy is an active research field, so improved approaches are proposed continuously; in this paper, we went with the standard implementation in the PySINDy package.

\section{Results}
\label{chapter:results}
To explore the performance of BPINNs, we assess them against two parameters in this paper: first, for varying levels of measurement noise in the training data, and, second, for varying lengths of the time series they receive as training input. 
 % two parameters of experimental evaluation were defined. 
The noise level in data $\mathcal{D}$ (measurement noise) follows an additive Gaussian noise model:
\begin{align}
    \eta = \mathcal{N}(0, \gamma_{noise})
    \label{eq:Noise_model}
\end{align}
with $\mathcal{N}$ being the Gaussian distribution of the measurement noise and $\gamma_{noise} = \bar{\boldsymbol{|x|}}\cdot K$, with $K \in [0.0, 0.05]$ and $\bar{\boldsymbol{|x|}}$ the mean absolute value of the state.This means that we examine noise levels that vary between 0\% and 5\%.
As a second parameter, we vary the length $T$ of the trajectories we use for training within the range $T\in[12 , 27] \,\text{s}$ and we investigate its influence on the estimation accuracy $\epsilon$. 

For both cases, we measure the performance of BPINNs against SINDy and PINNs using the percentage error (PE):
\begin{align}
    \boldsymbol{\epsilon} = \left|\frac{(\hat{\boldsymbol{\lambda}}-\boldsymbol{\lambda})}{\boldsymbol{\lambda}}\right| \cdot 100\,\% 
\end{align}
with $\hat{\boldsymbol{\lambda}}$ representing the mean of the corresponding posterior distribution (after 10 runs) in the case of BPINNs, and the mean of the point estimate (again, after 10 runs) in the case of SINDy and PINNs. 
%As mentioned earlier, all our results report for each of the methods the mean error over 10 runs, to avoid any effects related to initialization bias (due to the non-linear nature of the problem, different starting points can lead to slightly different performance).
For BPINNs, we also report the standard deviation $\boldsymbol{\Gamma}_{\text{post}}$ of the posterior distributions for the parameter estimates $\boldsymbol{\lambda}$, since BPINNs are able to deliver the standard deviation of their expected estimate in contrast to SINDy and PINNs. This can be interpreted as a measure of confidence for the estimates. In \cref{tbl:Noise_results} and \cref{tbl:trajectory_noise_resuts} we report the normalized values of $\boldsymbol{\Gamma}_{\text{post}}$, using the following equation:
\begin{align}
    \boldsymbol{\tau}=\boldsymbol{\Gamma}_{\text{post}}\cdot \frac{100\%}{\boldsymbol{\lambda}}
    \label{eq:tau}
\end{align}
The most certain estimate would give a standard deviation $\Gamma_{\text{post}}\approx 0$, and, therefore, $\tau \approx 0$. We expect $\epsilon$ and $\Gamma_{\text{post}}$ to increase with increasing noise in $\mathcal{D}$. At the same time, it is interesting to highlight that larger deviations in the training trajectories $\boldsymbol{x}$ should lead to smaller estimation errors $\epsilon$ and larger confidence (narrower $\Gamma_{\text{post}}$), as there are less functions found by the BPINN that fit $\mathcal{D}$. We will see the influence of the amount of deviations when we examine the trajectory length in Section~\ref{sec:traj_length} 

\subsection{Influence of noise}
\label{sec:noise}
To investigate the performance of BPINN under noise, a noise level ranging from $0\,\%$ to $5\,\%$ is added to each measurement, in steps of $1\,\%$ following \eqref{eq:Noise_model}. \cref{tbl:Noise_results} compares the resulting performance of the BPINN to the PINN and SINDy algorithm. 
%The results from the SINDy algorithm are shown as dots as they do not provide a distribution for their estimate. The parameter estimations from the BPINN are presented as blocks, confidence intervals, with the bar in the middle giving the mean, best guess, estimation. 
We observe that all three algorithms deliver a much more accurate estimate for the damping coefficient $d$ compared to the inertia constant $m$. Let us highlight here that BPINNs offer richer information than SINDy and PINNs through the standard deviation of the posterior distribution $\Gamma_{\text{post}}$ (presented in its normalized form $\tau$, see \eqref{eq:tau}). Through that, BPINNs allow us to understand that there is a much higher degree of certainty in the estimate of $d$, where $\tau$ is significantly smaller, compared to the estimate of $m$. The higher accuracy of the $d$ estimate relates to $d$ being primarily dependent on the frequency deviation $\Delta \omega$, while $m$ primarily depends on the rate of change of frequency $\dot{\Delta \omega}$ (ROCOF). Inspecting \cref{fig:trajectories_of_scenarios}, we observe that frequency deviation $\Delta \omega$ varies more than ROCOF, and, thus, the dataset contains more information useful for determining $d$. We continue this analysis in the discussion of the results, see \cref{sec:discussion}.

Comparing the performance of the three methods, we observe that for $d$, BPINN is not only the only one that offers a confidence measure, but also outperforms both SINDy and PINNs in the point estimate, achieving the lowest error $\epsilon_d$ across all noise levels and all fast/slow dynamics scenarios. This is not the case, however, for the estimate of $m$. In that case, PINNs appear to perform best for the fast dynamics, while SINDy performs best for the slow dynamics, as long as there is no noise. 

Besides the confidence measure that BPINNs offer, the most important result of this investigation is the robustness of both BPINNs and PINNs to noise. With noise gradually increasing from 0\% to 5\%, SINDy performance deteriorates rapidly, reaching even a 50\% estimation error. In contrast, BPINNs maintain an almost constant very low error for the estimate of $d$, independent of the noise level, while they also achieve a much lower error $\epsilon_m$ than SINDy for higher noise levels. PINNs seem to be equally robust to noise as BPINNs; and although they have higher errors than BPINNs for the estimate of $d$, they do achieve the lowest error across all scenarios for the estimate of $m$, with increasing noise levels.

\begin{table*}[!th]
\renewcommand{\arraystretch}{1.0}
\caption{Percent error (PE) $\epsilon$ and relative standard deviation $\tau$ for different noise levels}
\label{tbl:Noise_results}
\centering
\resizebox{\textwidth}{!}{%
\begin{tabular}{ll|llll|llll|llll|llll}
\toprule
$K$ & Algor. & \multicolumn{4}{c|}{Fast dynamics 1} & \multicolumn{4}{c|}{Fast dynamics 2} & \multicolumn{4}{c|}{Slow dynamics 1} & \multicolumn{4}{c}{Slow dynamics 2}\\
% &  & $\left|\frac{\hat{m} - m}{m}\right|$ [$\%$] & $\left|\frac{\hat{d} - d}{d}\right|$ [$\%$] & $\left|\frac{\hat{m} - m}{m}\right|$ [$\%$] & $\left|\frac{\hat{d} - d}{d}\right|$ [$\%$] &$\left|\frac{\hat{m} - m}{m}\right|$ [$\%$] & $\left|\frac{\hat{d} - d}{d}\right|$ [$\%$] & $\left|\frac{\hat{m} - m}{m}\right|$ [$\%$] & $\left|\frac{\hat{d} - d}{d}\right|$ [$\%$] \\
& & \multicolumn{2}{c}{Inertia Error} & \multicolumn{2}{c|}{Damping Error} & \multicolumn{2}{c}{Inertia Error} & \multicolumn{2}{c|}{Damping Error} & \multicolumn{2}{c}{Inertia Error} & \multicolumn{2}{c|}{Damping Error} & \multicolumn{2}{c}{Inertia Error} & \multicolumn{2}{c}{Damping Error} \\
& & $\epsilon_m$ [$\%$] & $\tau_m$ [$\%$] & $\epsilon_d$ [$\%$] & $\tau_d$ [$\%$] & $\epsilon_m$ [$\%$] & $\tau_m$ [$\%$] & $\epsilon_d$ [$\%$]  & $\tau_d$ [$\%$] & $\epsilon_m$ [$\%$] & $\tau_m$ [$\%$] & $\epsilon_d$ [$\%$] & $\tau_d$ [$\%$] & $\epsilon_m$ [$\%$]  & $\tau_m$ [$\%$] & $\epsilon_d$  [$\%$] & $\tau_d$ [$\%$] \\\midrule
\multirow{3}{*}{$0\%$} & BPINN & 1.20 & 9.26 & 0.01 & 0.011  & 2.45 & 28.03 & 0.01 & 0.005 & 2.38 & 53.55 & 0.002  & 0.001  & 5.69 & 53.54 & 0.001 & 0.002 \\

 & SINDy & 3.80 && 0.14 && 0.08 && 0.54  && 0.05 && 0.58 && 0.35 && 0.61&\\     
 &PINN & 0.34 && 0.84 && 0.19 && 1.22 && 1.08 && 1.41 && 3.16 && 1.15 & \\ \midrule
\multirow{3}{*}{$1\%$} & BPINN & 1.54 & 10.79 & 0.01 & 0.010 & 2.22 & 29.43  & 0.01 & 0.004 & 2.20 & 57.32 & 0.002 & 0.001 & 7.33 & 56.89 & 0.001 & 0.002 \\

 & SINDy & 14.76 && 1.38 && 22.82 && 0.75 && 4.49 && 8.37 && 5.72 && 5.41&\\     
 &PINN & 2.23 && 1.98 && 1.62 && 1.44 && 1.65 && 1.60 && 2.53 && 1.23 &\\    \midrule
\multirow{3}{*}{$2\%$} & BPINN & 1.50 & 11.78 & 0.01 & 0.011 & 2.06 & 38.47 & 0.01 & 0.005 & 3.53 & 61.13 & 0.002 & 0.001 & 10.53 & 64.14 & 0.013 & 0.002\\ 

 & SINDy & 13.21 && 4.13 && 44.43 && 1.45 && 28.93 && 24.89 && 9.16 &&  21.36 &\\         
 &PINN & 0.11 && 2.46 && 0.91 && 2.01 && 1.73 && 1.74 && 3.85 && 1.22 &\\    \midrule
\multirow{3}{*}{$3\%$} & BPINN & 1.83 & 25.82 & 0.02 & 0.010 & 3.54 & 47.62 & 0.01 & 0.005 & 2.94 & 81.93 & 0.003 & 0.001 & 11.44 & 67.55 & 0.016 & 0.004\\

 & SINDy & 19.54 && 2.18 && 21.40 && 3.07 && 24.86 && 34.06 && 20.86 && 32.71&\\        
 &PINN & 2.71 && 2.90 && 1.23 && 2.35 && 2.77 && 1.77 && 3.69 && 1.48 &\\    \midrule
 \multirow{3}{*}{$4\%$} & BPINN & 1.41 & 25.61 & 0.02 & 0.011 & 3.87 & 62.82 & 0.01 & 0.004 & 5.43 & 43.79 & 0.004 & 0.001 & 13.59 & 73.86 & 0.021 & 0.004\\
 
 & SINDy & 27.21 && 5.55 && 36.33 && 2.71 && 36.26 && 42.21 && 19.99 && 34.51 &\\   
 &PINN & 0.77 && 3.04 && 0.61 && 2.95 && 2.89 && 1.83 && 4.36 && 1.42 &\\     \midrule
 \multirow{3}{*}{$5\%$} & BPINN & 2.03 & 21.59 & 0.02 & 0.011 & 3.72 & 59.10 & 0.01 & 0.005 & 5.91 & 87.96 & 0.005 & 0.001 & 13.00 & 78.03 & 0.020 & 0.004\\
 
 & SINDy & 37.88 && 9.67 && 58.79 && 2.94 && 33.09 && 51.91 && 23.51 && 47.99 &\\       
 &PINN & 0.41 && 3.99 && 2.29 && 3.26 && 3.27 && 2.06 && 2.07 && 1.52 &\\    \bottomrule 
\end{tabular}%
}
\end{table*}

\subsection{Influence of trajectory length}
\label{sec:traj_length}
We expand our analysis, by including as parameters both the noise level and the length of the time series that the three methods receive as training input. \cref{tbl:trajectory_noise_resuts} presents the results for varying levels of noise $K$ between 0\% and 5\%, similar to \cref{sec:noise}, and varying trajectory length $T$ from the default $27\,\text{s}$ down to $12\,\text{s}$.

Here again, we observe the excellent performance of BPINNs for the estimate of $d$ across all noise levels and trajectory lengths, and the very good performance of PINNs for the estimate of $m$, for higher noise levels and all trajectory lengths. Here, it is interesting to highlight three phenomena. First, that SINDy appears to perform better for shorter trajectory lengths compared to the longer trajectory lengths in \cref{tbl:Noise_results}, although still an order of magnitude worse for higher noise, especially in systems with fast dynamics. Second, PINNs appear to have a consistent good performance with errors up to about 3\% across all scenarios. Third, BPINNs appear to perform significantly better in systems with faster dynamics compared to systems with slow dynamics. We will examine the reasons behind these phenomena in \cref{sec:discussion}.

% The influence of the length of the temporal sequence $T$ on the estimation error $\epsilon$ is now evaluated in \cref{tbl:trajectory_noise_resuts} together with different noise levels $K$. The estimation error of the BPINN, SINDy and PINN is shown for the four scenarios under the influence of different trajectory lengths in the range of $27\,\text{s}$ to $12\,\text{s}$ and noise levels as used before. 

% It can be seen that the BPINN estimations of $d$ show robustness against the noise and length of the trajectory in all evaluated scenarios and achieve comparable and better results compared to SINDy and PINN. For SINDy it can be seen, that the $\epsilon$ increases with decreasing $T$ and increasing $K$.
% When estimating parameter $m$ the BPINN estimations error significantly increases for decreasing trajectory length $T$ and increasing noise $K$ in slow dynamics. In contrast, SINDy estimation error increases the most for fast dynamics, while the PINN results show only small increase in $\epsilon$ for fast dynamics. 
% So, the BPINN estimation of $m$ seems to be sensitive to shorter trajectories in systems with slower dynamics. The reasons will be evaluated in the discussion section. It can be found again for increasing $\epsilon$ that the $\Gamma_{\text{post}}$ also increases, so the BPINN shows uncertainty quantification abilities. 

\begin{table*}[!th]
\renewcommand{\arraystretch}{0.8}
\caption{Percent error (PE) $\epsilon$ and relative standard deviation $\tau$ for different noise levels and trajectory length}
\label{tbl:trajectory_noise_resuts}
\centering
\resizebox{\textwidth}{!}{%
\begin{tabular}{ll|llll|llll|llll|llll}
\toprule
 & Algor. & \multicolumn{4}{c|}{Fast dynamics 1} & \multicolumn{4}{c|}{Fast dynamics 2} & \multicolumn{4}{c|}{Slow dynamics 1} & \multicolumn{4}{c}{Slow dynamics 2}\\
% &  & $\left|\frac{\hat{m} - m}{m}\right|$ [$\%$] & $\left|\frac{\hat{d} - d}{d}\right|$ [$\%$] & $\left|\frac{\hat{m} - m}{m}\right|$ [$\%$] & $\left|\frac{\hat{d} - d}{d}\right|$ [$\%$] &$\left|\frac{\hat{m} - m}{m}\right|$ [$\%$] & $\left|\frac{\hat{d} - d}{d}\right|$ [$\%$] & $\left|\frac{\hat{m} - m}{m}\right|$ [$\%$] & $\left|\frac{\hat{d} - d}{d}\right|$ [$\%$] \\
($K$; & & \multicolumn{2}{c}{Inertia Error} & \multicolumn{2}{c|}{Damping Error} & \multicolumn{2}{c}{Inertia Error} & \multicolumn{2}{c|}{Damping Error} & \multicolumn{2}{c}{Inertia Error} & \multicolumn{2}{c|}{Damping Error} & \multicolumn{2}{c}{Inertia Error} & \multicolumn{2}{c}{Damping Error} \\
 $T$) & & $\epsilon_m$ [$\%$] & $\tau_m$ [$\%$] & $\epsilon_d$ [$\%$] & $\tau_d$ [$\%$] & $\epsilon_m$ [$\%$] & $\tau_m$ [$\%$] & $\epsilon_d$ [$\%$]  & $\tau_d$ [$\%$] & $\epsilon_m$ [$\%$] & $\tau_m$ [$\%$] & $\epsilon_d$ [$\%$] & $\tau_d$ [$\%$] & $\epsilon_m$ [$\%$]  & $\tau_m$ [$\%$] & $\epsilon_d$  [$\%$] & $\tau_d$ [$\%$] \\\midrule
\multirow{3}{*}{$(0\%;$} & BPINN & 1.20 & 9.26  & 0.01 & 0.011  & 2.45 & 28.03 & 0.01 & 0.005 & 2.38 & 53.55 & 0.002 & 0.001 & 5.69 & 53.54 & 0.001 & 0.002 \\

 & SINDy & 3.80 & & 0.14& & 0.08& & 0.54&  & 0.05& & 0.58& & 0.35 && 0.61&\\      
$27\,\text{s})$ & PINN & 0.34 && 0.84& & 0.19& & 1.22 & & 1.08 && 1.41 && 3.16 && 1.15&\\   \midrule
\multirow{3}{*}{$(1\%;$} & BPINN & 1.88 & 6.81 & 0.01 &  0.011 & 1.71 & 28.50 & 0.01 & 0.004 & 3.14 & 55.40 & 0.002 & 0.001 & 9.38 & 55.62 & 0.01 & 0.002\\  

 & SINDy & 1.87 && 0.50 && 22.69 && 0.34 && 2.06 && 3.24 && 1.43 && 2.37&\\         
$24\,\text{s})$ & PINN & 1.13 && 1.29 && 0.53 && 0.535 && 2.01 && 1.56 && 2.01 && 1.28&\\    \midrule
\multirow{3}{*}{$(2\%;$} & BPINN & 1.19 & 4.99 & 0.02 & 0.011 & 3.28 & 41.99 & 0.02 & 0.005 & 4.51 & 61.88 & 0.004 & 0.001 & 13.38 & 62.67 & 0.02 & 0.005\\

 & SINDy & 2.31 && 1.17 && 11.66 && 0.82 && 4.01 && 7.97 && 2.02 && 5.78&\\         
$21\,\text{s})$ & PINN & 1.96 && 1.27 && 0.32 && 1.58 && 0.92 && 1.52 && 1.08 && 1.24& \\    \midrule
\multirow{3}{*}{$(3\%;$} & BPINN & 2.18 & 12.89 & 0.02 & 0.011 & 2.55 & 36.85 & 0.02 & 0.004 & 7.80 & 73.11 & 0.005 & 0.001 & 19.70 & 65.94 & 0.03 & 0.006\\

 & SINDy & 9.53 && 0.88  && 37.26 && 1.46 && 8.78 && 11.27 && 4.45 && 8.39& \\
$18\,\text{s})$ & PINN & 0.23 && 1.18  && 0.83 && 1.72 && 1.14 && 1.72 && 1.84 && 1.47&\\    \midrule
 \multirow{3}{*}{$(4\%;$} & BPINN & 2.16 & 6.24 & 0.03 & 0.011 & 1.23 & 31.41 & 0.02 & 0.005 & 9.38 & 81.09 & 0.05 & 0.006 & 19.57 & 78.59 & 0.02 & 0.017\\
 
 & SINDy & 27.09 && 3.41 && 31.19 && 1.34 && 11.50 && 12.41 && 9.54 && 8.56&\\
$15\,\text{s})$ & PINN & 0.71 && 1.42 && 0.32 && 2.72 && 2.21 && 2.43 && 2.42 && 2.20& \\        \midrule
 \multirow{3}{*}{$(5\%;$} & BPINN & 2.17 & 2.27 & 0.04 & 0.009 & 2.25 & 42.89 & 0.02 & 0.008 & 6.84 & 105.98 & 0.01 & 0.036 & 16.57 & 108.09 & 0.01 & 0.092\\
 
 & SINDy & 41.30 && 12.56 && 42.89 && 2.09 && 10.84 && 10.05 && 8.87 && 6.17&\\
$12\,\text{s})$ & PINN & 2.04 && 2.24 && 1.58 && 1.40 && 2.85 && 1.65 && 3.05 && 1.45 &\\        \bottomrule 
\end{tabular}%
}
\end{table*}

\subsection{Discussion}
\label{sec:discussion}

\paragraph{BPINN vs SINDy Performance under noisy data} From \cref{tbl:Noise_results} and \cref{tbl:trajectory_noise_resuts} it is clear that in the presence of noise, the BPINN provides estimates that are significantly better than SINDy, and comparable to PINN. The reason for this can be found in the differentiation methods. SINDy is calculating the derivative for every point in $\mathcal{D}$ and tries to find suitable parameters for the resulting differential equations. So, in high noise, the SINDy derivatives can have large errors between the single points. This is different for the BPINN and PINN, because they fit the whole trajectory $\hat{\boldsymbol{x}}$ first, and then calculate the derivative of the resulting fit $\frac{d}{dt}\hat{\boldsymbol{x}}$.

\begin{figure}[h]
    \centering
    \includegraphics[clip, width=3.5in]{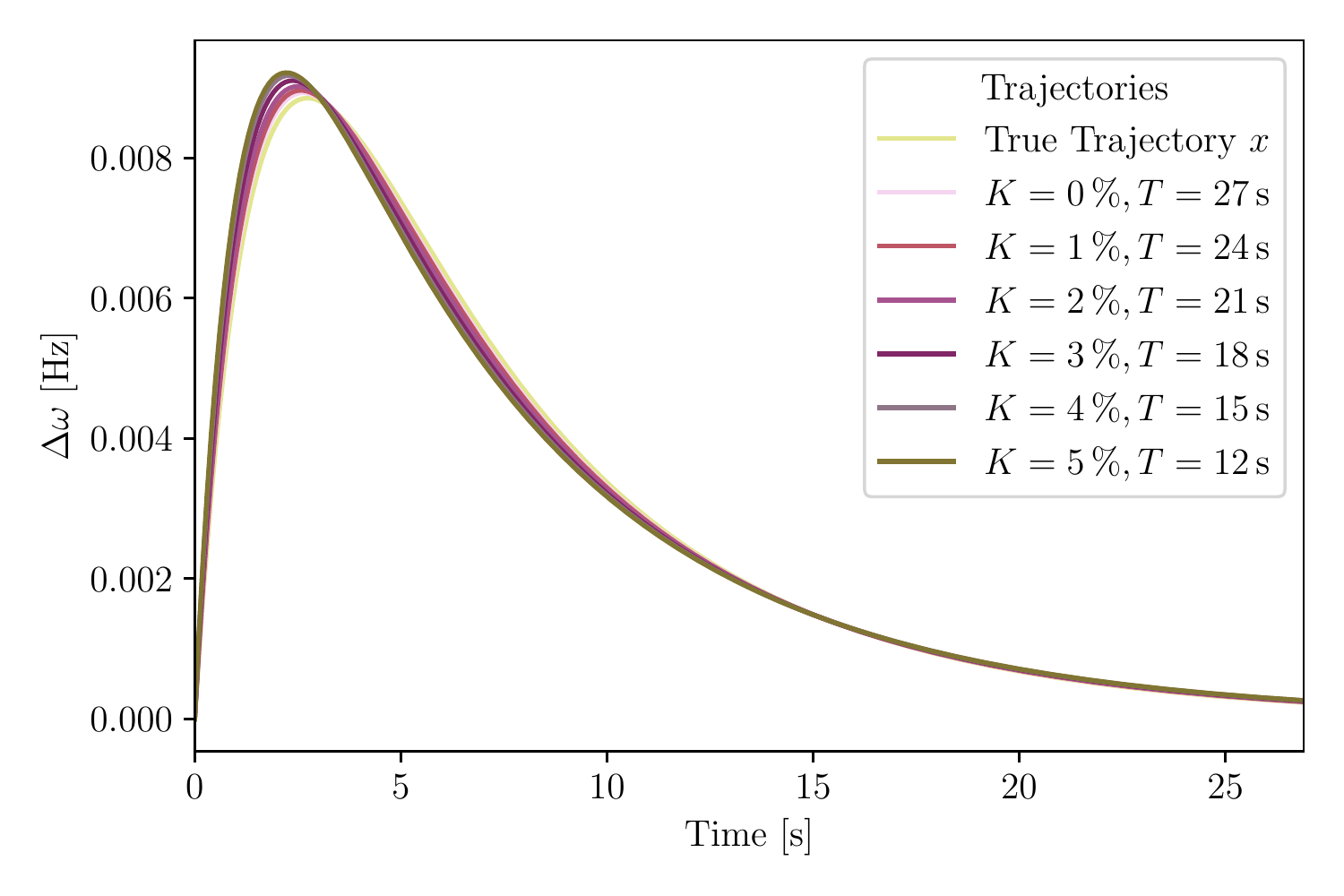}
    \caption{Reconstruction of $\Delta \omega$ with BPINN $m$ estimates for the Slow Dynamics 2 scenario from \cref{tbl:trajectory_noise_resuts}}
    \label{fig:reconst_m_ld}
\end{figure}

\paragraph{BPINN Performance -- $m$ vs $d$} BPINNs appear to achieve 2 to 3 orders of magnitude lower estimation errors for $d$ compared to $m$. This is more pronounced when estimating $m$ and $d$ in slow dynamics, and when using shorter trajectory lengths. Indeed, in slow dynamics, the trajectory of $\boldsymbol{x}$ contains significantly less information about $m$. To investigate the underlying reasons, in \cref{fig:reconst_m_ld} we plot the reconstruction of the $\Delta \omega$ trajectory for all input trajectory lengths $T$ of the Slow Dynamics 2 scenario, using the BPINN estimates found in \cref{tbl:trajectory_noise_resuts}. 
% as illustrated in the reconstruction plot in \cref{fig:reconst_m_ld}. In this Figure, we present . 
%The small influence of $m$ has huge influence on the BPINN estimation error and confidence interval width. 
We observe that the resulting trajectory is not sensitive to varying parameter $m$: despite considering BPINN estimates which had a non-negligible estimation error $\epsilon_m$, all trajectories look similar and closely overlap with the true trajectory $x$. This also shows that systems with higher inertia and damping are less sensitive to larger variations of the inertia constant $m$. This, however, also results to larger estimation errors, as the BPINN can fit a wider range of possible functions from the family of functions to the training data. % in the BPINN. 
Same goes for $\boldsymbol{\Gamma}_{post}$, as a larger $\boldsymbol{\Gamma}_{post}$ does not result in significantly increased loss during training, since these functions do not contradict the target trajectory. This is also the reason that PINN performs better in these cases: PINNs have to fit only a single function and not a family of functions as the BPINN does.
Similar is the case when we consider shorter trajectory lengths for training. Assuming the same sampling frequency for all lengths $T$, there are fewer points that contradict the trajectory estimate of the BPINN. So, the loss during training is small for a wider range of functions, and this makes them appear as a suitable fit for the optimizer. This is the reason that for lower noise, both SINDy and PINN achieve more accurate estimates of $m$. Still, we shall mention that for cases with slower dynamics, larger estimation errors of $m$ similar to BPINNs do not seem to significantly affect the expected dynamic performance, as shown in \cref{fig:reconst_m_ld}.
%When experiencing slower dynamics, both algorithms SINDy and the proposed BPINN parameter estimates are usually less accurate. This could be explained by the amount of information that is found in the provided temporal sequence. As the length of the trajectory is the same, the full swing is not in the range of simulation in scenario $2$ and $3$ and also the deviation of the frequency and angle are not as large over the temporal sequence as shown in \cref{fig:trajectories_of_scenarios}.\\
%It has been shown that the BPINN is generally able to perform well for different dynamics of the given problem under noise influence and also different trajectory lengths and achieves comparable results to the SINDy algorithm. The advantage of the BPINN is being able to give a confidence about its estimate which provides the user valuable information and showing robustness against noise compared to SINDy.
\paragraph{Confidence measure} We claim that the magnitude of the normalized standard deviation $\tau$ not only gives a form parameter of the posterior distribution but it also contains information about how `confident' the BPINN is that the delivered point estimate, i.e. mean of posterior distribution, is accurate. In \cref{fig:std_over_mean_err} we plot the normalized standard deviation $\tau$ of the posterior distributions against the error of the mean estimate $\epsilon$ for $m$ and $d$, for the four scenarios. Indeed, we observe a clear correlation, where the standard deviation is monotonously increasing with increasing estimation error for both parameters. Such a confidence measure can be very valuable for the operation of safety critical systems, such as power systems, as the magnitude of the posterior distributions standard deviation can give a useful indication about how much we can trust the point estimate of the BPINN about e.g. the system inertia. Methods such as conventional neural networks or SINDy do not offer any information about how `trustworthy' is their output. Future work shall perform detailed correlation analysis utilizing a wide range of scenarios. 
% It should be highlighted here, that this measure of confidence can be very valuable, especially in operation of safety critical systems, e.g. power systems.  
%\begin{figure}[h]
%    \centering
%    \includegraphics[clip, width=3.5in]{std_over_err_m.pdf}
%    \caption{Standard deviation of all $m$ estimates over mean err}
%    \label{fig:std_over_mean_err}
%\end{figure}

\begin{figure}
    \renewcommand{\arraystretch}{0.3}
    \centering
    \begin{subfigure}{0.48\linewidth}
        \includegraphics[width=\linewidth]{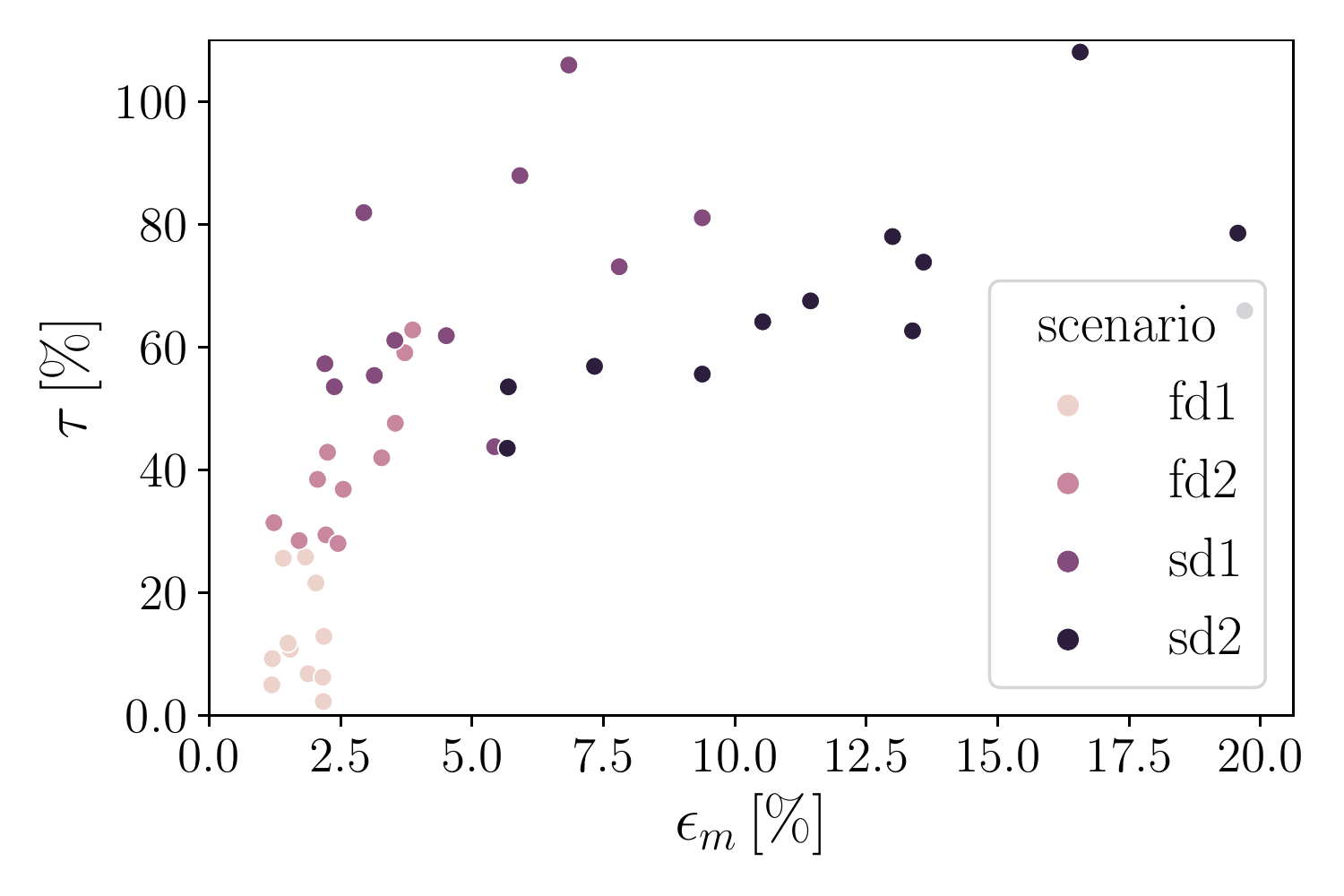}
        \caption{Standard deviation of all $m$ estimates versus the error}
        \label{fig:std_over_mean_err}
    \end{subfigure}
    \hfill
    \begin{subfigure}{0.48\linewidth}
        \includegraphics[width=\linewidth]{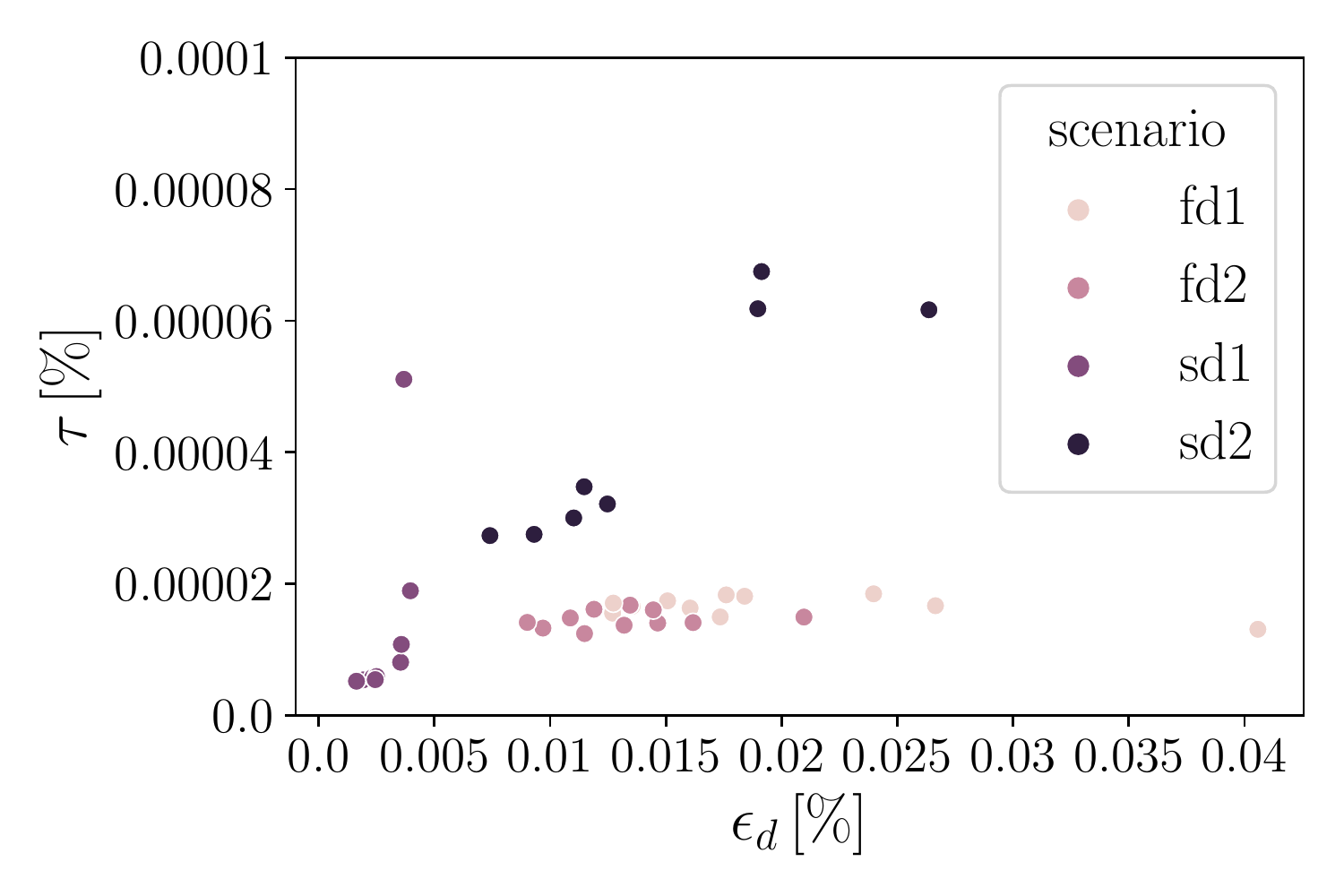}
        \caption{Standard deviation of all $d$ estimates versus the error}
        \label{fig:std_over_mean_err}
    \end{subfigure}
    \caption{Standard deviation $\tau$ versus estimate errors $\epsilon$}
    \label{fig:std_over_mean_err}
\end{figure}

\paragraph{Initialization of the priors} Although BPINNs do not require prior knowledge about the target parameters or the noise, they do require the initialization of the physical priors $\boldsymbol{\lambda}$. It appears that this initialization should be chosen carefully to achieve small errors $\epsilon$ when the noise in $\mathcal{D}$ is unknown, which would be the case when receiving measurement data. So, a general prior that works for all noise levels and all scenarios should be determined.
We found through parameter studies that a normal distribution $P(\boldsymbol{\lambda})$ can be used as follows 
\begin{align}
    P(\boldsymbol{\lambda}) &\sim \mathcal{N}(1.0, 5.0/P_{prec})\\
    P_{prec} &\sim \Gamma(1.0, 0.1)
\end{align}
where $P(\boldsymbol{\lambda})$ depends on a precision Gamma distribution of $P_{prec}$ with shape parameter $\alpha=1$ and rate parameter $\beta=0.1$. The resulting probability density function, besides being as uninformative as possible due to the $\Gamma$ dependency, also limits the samples to being positive. This is valid, since negative values for $\boldsymbol{\lambda}=\{m,d\}$ are physically impossible. It also provides a prior, that avoids a biased system, while providing the expected shape (Gaussian), so the estimation result does not rely on previously available knowledge, which the prior is normally utilized for.

\paragraph{Runtime} The average runtime of SINDy was $0.0032\,\text{s}$ for one estimate. The BPINN training took on average $80\,\text{s}$ while PINN training took $18.53\,\text{s}$. All methods were trained on an Intel i7 CPU. The increased computation time for the BPINN can be explained by the family of functions that has to be fit in comparison to the PINN, which needs to fit a neural network to only a single function. Still, there is significant potential to decrease the computation time for the BPINN with the utilization of GPU computing and code optimization, which is object of our future work.

\section{Conclusion and Outlook}
\label{chapter: conclusion}
This paper, to the best of our knowledge, introduced for the first time the Bayesian Physics-Informed Neural Networks for applications in power systems. In contrast to existing approaches, Bayesian Physics-Informed Neural Networks (BPINNs) deliver a confidence measure of their estimates, while they do not require inaccurate assumptions about the prior distributions of the variables to be determined. 

This paper applies the BPINNs for robust identification of the system inertia and damping, using a single machine infinite bus system as the guiding example. Comparing BPINNs with the Physics-Informed Neural Networks (PINNs) and the recently popular approach for System Identification, SINDy, we assess their performance. We find that BPINNs and PINNs are robust against all noise levels, delivering estimates of the system inertia and damping with significantly lower error compared to SINDy, especially as the noise levels increased.
% On the contrary, increasing noise in the input measurements results in a significantly poorer performance of SINDy. 

% On the other hand, BPINNs may suffer if the input data contain limited information about the system parameters. In our tests, slow dynamic scenarios along with shorter lengths of the input trajectory contain lower information about the system inertia, and this led BPINNs to less accurate estimates; in contrast, SINDy achieved a lower error in these cases. The opposite occurs in the case of fast dynamics. In almost all fast cases, and especially if there were noisy data, BPINNs outperformed SINDy by one order of magnitude.  

% It is also worth highlighting that conventional PINNs achieved a consistently good performance across all scenarios. In contrast, BPINNs offer a confidence measure of their estimate, which can be extremely useful for building the necessary trust for the application of Neural Networks in general, and BPINNs in particular, in safety-critical systems, such as power systems. On the other hand, methods such as conventional neural networks or SINDy do not offer any information about how `trustworthy' is their output. 

% This paper attempts a first effort on the application of Bayesian Physics Informed Neural Networks in power systems, applying them on robust system identification.
BPINNs combine the strengths of the Physics-Informed Neural Networks with a confidence measure for their outputs, delivering a tool that can offer a degree of `trustworthiness' of their output. 
% This is especially relevant for safety-critical systems, such as power systems. 
This opens up a wide range of opportunities for applications in power systems, that remain to be explored.

% about the  prior knowledge about the parameters themselves

% This approach has uncertainty quantification abilities and provides a confidence measure about its estimate through the standard deviation of the posterior distribution of the physical parameters. Furthermore, it allows to perform the estimation without any prior knowledge about the parameters itself or the region in which the parameters should presumably be.
% The approach is evaluated utilizing a single-machine-infinite-bus system, which can be seen as the representation of part of a real system presented as an aggregated generator. For comparison, the PINN and SINDy algorithm are used. In future studies, the approach should be evaluated in larger systems, estimating the system parameters at multiple points in the grid, and should also be applied to other problems in power systems, as it allows solving inverse problems when there is general knowledge about the system equations. 

\bibliographystyle{ieeetr}
\bibliography{library.bib}

\end{document}